\documentstyle[12pt]{article}
\newcommand{\rl}{\rule[-0.6cm]{0cm}{1.2cm}}
\newcommand{\rll}{\rule[-0.3cm]{0cm}{1.cm}}
\newcommand{\sts}{\footnotesize}
\newcommand{\scr}{\scriptsize}

\newcommand{\spz}{\hspace{0.7cm}}
\newcommand{\sps}{\hspace{3mm}}
\newcommand{\stl}{\small}
\newcommand{\nn}{\nonumber}
\newcommand{\fr}{\rightarrow}
\newcommand{\de}{\partial}
\newcommand{\ri}{\right}
\newcommand{\lf}{\left}
\newcommand{\th}{\theta}

\newcommand{\Th}{\Theta}
\newcommand{\ep}{\varepsilon}
\newcommand{\eq}{\begin{equation}}
\newcommand{\en}{\end{equation}}
\newcommand{\bea}{\begin{eqnarray}}
\newcommand{\eea}{\end{eqnarray}}
\newcommand{\acc}{\\[3mm]}
\newcommand{\ba}{\begin{array}}
\newcommand{\ea}{\end{array}}
\newcommand{\ds}{\displaystyle}
\newcommand{\virg}{\spz,}
\newcommand{\pu}{\spz.}
\newcommand{\resection}[1]{\setcounter{equation}{0}\section{#1}}

\newcommand{\Z}{{\cal Z}}
\newcommand{\SO}{{\cal SO}}

\newcommand{\SU}{{\cal SU}}
\newcommand{\M}{{\cal M}}
\newcommand{\A}{{\cal A}}
\newcommand{\D}{{\cal D}}
\newcommand{\E}{{\cal E}}
\newcommand{\T}{{\cal T}}
\newcommand{\G}{{\cal G}}
\newcommand{\F}{{\cal F}}
\newcommand{\B}{{\cal B}}
\newcommand{\C}{{\cal C}}

\newcommand{\NP}[1]{Nucl.\ Phys.\ {\bf #1}}
\newcommand{\PL}[1]{Phys.\ Lett.\ {\bf #1}}

\newcommand{\CMP}[1]{Comm.\ Math.\ Phys.\ {\bf #1}}

\newcommand{\MPL}[1]{Mod.\ Phys.\ Lett.\ {\bf #1}}
\newcommand{\IJMP}[1]{Int.\ J.\ Mod.\ Phys.\ {\bf #1}}

\begin{document}
\setlength{\unitlength}{.8mm}
\newsavebox{\Bn}
\sbox{\Bn}{\begin{picture}(52,5)(0,-3.5)
\multiput(10,0)(10,0){5}{\circle*{3}}
\multiput(10,0)(10,0){1}{\line(1,0){10}}
\multiput(21,0)(1,0){9}{\circle*{.2}}
\put(30,0){\line(1,0){10}}
\put(40,0.5){\line(1,0){10}}
\put(40,-0.5){\line(1,0){10}}
\put(12,-2){\makebox(0,0)[t]{{\protect\scr 1}}}
\put(22,-2){\makebox(0,0)[t]{{\protect\scr 2}}}
\put(32,-2){\makebox(0,0)[t]{{\protect\scr {\em n}--2}}}
\put(42,-2){\makebox(0,0)[t]{{\protect\scr {\em n}--1}}}
\put(52,-2){\makebox(0,0)[t]{{\protect\scr {\em n}}}}
\end{picture}}
\newsavebox{\BnTBA}
\sbox{\BnTBA}{\begin{picture}(52,5)(0,-3.5)
\multiput(10,0)(10,0){5}{\circle*{3}}
\multiput(10,0)(10,0){1}{\line(1,0){10}}
\multiput(21,0)(1,0){9}{\circle*{.2}}
\put(30,0){\line(1,0){10}}
\put(40,0.5){\line(1,0){10}}
\put(40,-0.5){\line(1,0){10}}
\put(50,1.5){\line(0,1){7}}
\put(50,-1.5){\line(0,-1){7}}
\put(50,10){\circle{3}}
\put(50,-10){\circle{3}}
\put(56,10){\makebox(0,0)[t]{{\protect\scr ({\em n},3)}}}
\put(56,-10){\makebox(0,0)[t]{{\protect\scr ({\em n},1)}}}
\put(12,-2){\makebox(0,0)[t]{{\protect\scr 1}}}
\put(22,-2){\makebox(0,0)[t]{{\protect\scr 2}}}
\put(32,-2){\makebox(0,0)[t]{{\protect\scr {\em n}--2}}}
\put(42,-2){\makebox(0,0)[t]{{\protect\scr {\em n}--1}}}
\put(56,-2){\makebox(0,0)[t]{{\protect\scr ({\em n},2)}}}
\end{picture}}
\newsavebox{\pole}
\sbox{\pole}
{\begin{picture}(2,2)(0,0)
\put(0,0){\line(1,3){1.5}}
\put(0,0){\line(-1,3){1.5}}
\put(0,4.5){\oval(3,3)[t]}
\end{picture}}
\newsavebox{\Tn}
\sbox{\Tn}{\begin{picture}(55,10)(0,-3.5)
\multiput(10,0)(10,0){5}{\circle*{3}}
\multiput(10,0)(10,0){2}{\line(1,0){10}}
\multiput(31,0)(1,0){9}{\circle*{.2}}
\put(40,0){\line(1,0){10}}
\put(50,1.2){\usebox{\pole}}
\put(12,-2){\makebox(0,0)[t]{{\protect\scr 1}}}
\put(22,-2){\makebox(0,0)[t]{{\protect\scr 2}}}
\put(32,-2){\makebox(0,0)[t]{{\protect\scr 3}}}
\put(42,-2){\makebox(0,0)[t]{{\protect\scr {\em n}--2}}}
\put(52,-2){\makebox(0,0)[t]{{\protect\scr {\em n}--1}}}
\end{picture}}
\newsavebox{\Cn}
\sbox{\Cn}{\begin{picture}(52,5)(0,-3.5)
\multiput(10,0)(10,0){5}{\circle*{3}}
\multiput(10,0)(10,0){1}{\line(1,0){10}}
\multiput(21,0)(1,0){9}{\circle*{.2}}
\put(30,0){\line(1,0){10}}
\put(40,0.5){\line(1,0){10}}
\put(40,-0.5){\line(1,0){10}}
\put(12,-2){\makebox(0,0)[t]{{\protect\scr 1}}}
\put(22,-2){\makebox(0,0)[t]{{\protect\scr 2}}}
\put(32,-2){\makebox(0,0)[t]{{\protect\scr {\em n}--2}}}
\put(42,-2){\makebox(0,0)[t]{{\protect\scr {\em n}--1}}}
\put(52,-2){\makebox(0,0)[t]{{\protect\scr {\em n}}}}
\end{picture}}
\newsavebox{\CnTBA}
\sbox{\CnTBA}{\begin{picture}(52,5)(0,-3.5)
\multiput(10,0)(10,0){5}{\circle*{3}}
\multiput(10,1.5)(10,0){4}{\line(0,1){7}}
\multiput(10,-1.5)(10,0){4}{\line(0,-1){7}}
\multiput(10,-10)(10,0){4}{\circle{3}}
\multiput(10,10)(10,0){4}{\circle{3}}
\multiput(10,0)(10,0){1}{\line(1,0){10}}
\multiput(21,0)(1,0){9}{\circle*{.2}}
\put(30,0){\line(1,0){10}}
\put(40,0.5){\line(1,0){10}}
\put(40,-0.5){\line(1,0){10}}
\put(14.5,-2){\makebox(0,0)[t]{{\protect\scr (1,2)}}}
\put(45,-3){\makebox(0,0)[t]{{\protect\scr ({\em n}-1,2)}}}
\put(8.5,-13){\makebox(0,0)[t]{{\protect\scr (1,1)}}}
\put(49.5,-13){\makebox(0,0)[t]{{\protect\scr ({\em n}-1,1)}}}
\put(8.5,+16){\makebox(0,0)[t]{{\protect\scr (1,3)}}}
\put(48.5,+14){\makebox(0,0)[t]{{\protect\scr ({\em n}-1,3)}}}
\put(54,0){\makebox(0,0)[t]{{\protect\scr {\em n}}}}
\end{picture}}
\newsavebox{\FF}
\sbox{\FF}{\begin{picture}(52,5)(0,-3.5)
\multiput(10,0)(10,0){4}{\circle*{3}}
\put(10,0){\line(1,0){10}}
\put(30,0){\line(1,0){10}}
\put(20,0.5){\line(1,0){10}}
\put(20,-0.5){\line(1,0){10}}
\put(12,-2){\makebox(0,0)[t]{{\protect\scr 1}}}
\put(22,-2){\makebox(0,0)[t]{{\protect\scr 2}}}
\put(32,-2){\makebox(0,0)[t]{{\protect\scr 3}}}
\put(42,-2){\makebox(0,0)[t]{{\protect\scr 4}}}
\end{picture}}
\newsavebox{\FFTBA}
\sbox{\FFTBA}{\begin{picture}(52,5)(0,-3.5)
\multiput(10,0)(10,0){4}{\circle*{3}}
\put(10,0){\line(1,0){10}}
\put(20,-0.5){\line(1,0){10}}
\put(20,+0.5){\line(1,0){10}}
\put(30,0){\line(1,0){10}}
\multiput(30,1.5)(10,0){2}{\line(0,1){7}}
\multiput(30,-1.5)(10,0){2}{\line(0,-1){7}}
\multiput(30,-10)(10,0){2}{\circle{3}}
\multiput(30,10)(10,0){2}{\circle{3}}
\put(12,-2){\makebox(0,0)[t]{{\protect\scr 1}}}
\put(22,-2){\makebox(0,0)[t]{{\protect\scr 2}}}
\put(35,-2){\makebox(0,0)[t]{{\protect\scr (3,2)}}}
\put(45,-2){\makebox(0,0)[t]{{\protect\scr (4,2)}}}
\put(32,-13){\makebox(0,0)[t]{{\protect\scr (3,1)}}}
\put(42,-13){\makebox(0,0)[t]{{\protect\scr (4,1)}}}
\put(32,+16){\makebox(0,0)[t]{{\protect\scr (3,3)}}}
\put(42,+16){\makebox(0,0)[t]{{\protect\scr (4,3)}}}
\end{picture}}
\newsavebox{\GG}
\sbox{\GG}{\begin{picture}(52,5)(0,-3.5)
\multiput(10,0)(10,0){2}{\circle*{3}}
\put(10,0){\line(1,0){10}}
\put(10,+0.7){\line(1,0){10}}
\put(10,-0.7){\line(1,0){10}}
\put(12,-2){\makebox(0,0)[t]{{\protect\scr 1}}}
\put(22,-2){\makebox(0,0)[t]{{\protect\scr 2}}}
\end{picture}}
\newsavebox{\GGTBA}
\sbox{\GGTBA}{\begin{picture}(52,5)(0,-3.5)
\multiput(10,0)(10,0){2}{\circle*{3}}
\put(10,0){\line(1,0){10}}
\put(10,+0.7){\line(1,0){10}}
\put(10,-0.7){\line(1,0){10}}
\put(20,1.5){\line(0,1){7}}
\put(20,10){\circle{3}}
\put(26,10){\makebox(0,0)[t]{{\protect\scr (2,4)}}}
\put(20,-1.5){\line(0,-1){7}}
\put(20,-10){\circle{3}}
\put(26,-10){\makebox(0,0)[t]{{\protect\scr (2,2)}}}
\put(20,11.5){\line(0,1){7}}
\put(20,20){\circle{3}}
\put(26,20){\makebox(0,0)[t]{{\protect\scr (2,5)}}}
\put(20,-11.5){\line(0,-1){7}}
\put(20,-20){\circle{3}}
\put(26,-20){\makebox(0,0)[t]{{\protect\scr (2,1)}}}
\put(12,-2){\makebox(0,0)[t]{{\protect\scr 1}}}
\put(26,-2){\makebox(0,0)[t]{{\protect\scr (2,3)}}}
\end{picture}}
\newsavebox{\SBnTBA}
\sbox{\SBnTBA}{\begin{picture}(52,5)(0,-3.5)
\multiput(10,0)(10,0){5}{\circle{3}}
\multiput(10,0)(10,0){1}{ \line(1,0){7} }
\multiput(21.5,0)(1,0){8}{\circle*{.2}}
\multiput(31.5,0)(1,0){8}{\circle*{.2}}
\put(41.5,0.5){\line(1,0){7}}
\put(41.5,-0.5){\line(1,0){7}}
\put(50,1.5){\line(0,1){7}}
\put(50,-1.5){\line(0,-1){7}}
\put(50,10){\circle{3}}
\put(50,-10){\circle{3}}
\put(12,-2){\makebox(0,0)[t]{{\protect\scr 1}}}
\put(22,-2){\makebox(0,0)[t]{{\protect\scr 2}}}
\put(32,-2){\makebox(0,0)[t]{{\protect\scr {\em k}}}}
\put(30,0){\circle*{3}}
\put(42,-2){\makebox(0,0)[t]{{\protect\scr {\em n}--1}}}
\put(56,10){\makebox(0,0)[t]{{\protect\scr  ({\em n},3)}}}
\put(56,-10){\makebox(0,0)[t]{{\protect\scr ({\em n},1)}}}
\put(56,-2){\makebox(0,0)[t]{{\protect\scr  ({\em n},2)}}}
\end{picture}}
\newsavebox{\SBBnTBA}
\sbox{\SBBnTBA}{\begin{picture}(52,5)(0,-3.5)
\multiput(10,0)(10,0){5}{\circle{3}}
\multiput(10,0)(10,0){1}{ \line(1,0){7} }
\multiput(21.5,0)(1,0){8}{\circle*{.2}}
\multiput(31.5,0)(1,0){8}{\circle*{.2}}
\put(41.5,0.5){\line(1,0){7}}
\put(41.5,-0.5){\line(1,0){7}}
\put(50,1.5){\line(0,1){7}}
\put(50,-1.5){\line(0,-1){7}}
\put(50,0){\circle*{3}}
\put(50,10){\circle*{3}}
\put(50,-10){\circle*{3}}
\put(12,-2){\makebox(0,0)[t]{{\protect\scr 1}}}
\put(22,-2){\makebox(0,0)[t]{{\protect\scr 2}}}
\put(42,-2){\makebox(0,0)[t]{{\protect\scr {\em n}--1}}}
\put(56,10){\makebox(0,0)[t]{{\protect\scr  ({\em n},3)}}}
\put(56,-10){\makebox(0,0)[t]{{\protect\scr ({\em n},1)}}}
\put(56,-2){\makebox(0,0)[t]{{\protect\scr ({\em n},2)}}}
\end{picture}}
\newsavebox{\SDnTBA}
\sbox{\SDnTBA}{\begin{picture}(52,5)(0,-3.5)
\multiput(10,0)(10,0){5}{\circle{3}}
\multiput(11.5,0)(10,0){1}{\line(1,0){7}}
\multiput(21.5,0)(1,0){8}{\circle*{.2}}
\multiput(31.5,0)(1,0){8}{\circle*{.2}}
\put(41.5,0){\line(1,0){7}}
\put(51,1){\line(1,1){8}}
\put(51,-1){\line(1,-1){8}}
\put(60,10){\circle{3}}
\put(60,-10){\circle{3}}
\put(68,10){\makebox(0,0)[t]{{\protect\scr {\em n}-1}}}
\put(68,-10){\makebox(0,0)[t]{{\protect\scr {\em n}}}}
\put(12,-2){\makebox(0,0)[t]{{\protect\scr 1}}}
\put(22,-2){\makebox(0,0)[t]{{\protect\scr 2}}}
\put(32,-2){\makebox(0,0)[t]{{\protect\scr {\em k}}}}
\put(30,0){\circle*{3}}
\put(42,-2){\makebox(0,0)[t]{{\protect\scr {\em n}--3}}}
\put(48,-2){\makebox(0,0)[t]{{\protect\scr {\em n}--2}}}
\end{picture}}
\newsavebox{\SDDnTBA}
\sbox{\SDDnTBA}{\begin{picture}(52,5)(0,-3.5)
\multiput(10,0)(10,0){5}{\circle{3}}
\multiput(11.5,0)(10,0){1}{\line(1,0){7}}
\multiput(21.5,0)(1,0){8}{\circle*{.2}}
\multiput(31.5,0)(1,0){8}{\circle*{.2}}
\put(41.5,0){\line(1,0){7}}
\put(51,1){\line(1,1){8}}
\put(51,-1){\line(1,-1){8}}
\put(60,10){\circle*{3}}
\put(60,-10){\circle*{3}}
\put(68,10){\makebox(0,0)[t]{{\protect\scr {\em n}-1}}}
\put(68,-10){\makebox(0,0)[t]{{\protect\scr {\em n} }}}
\put(12,-2){\makebox(0,0)[t]{{\protect\scr 1}}}
\put(22,-2){\makebox(0,0)[t]{{\protect\scr 2}}}
\put(42,-2){\makebox(0,0)[t]{{\protect\scr {\em n}--3}}}
\put(48,-2){\makebox(0,0)[t]{{\protect\scr {\em n}--2}}}
\end{picture}}
\newsavebox{\STNTBA}
\sbox{\STNTBA}{\begin{picture}(55,10)(0,-3.5)
\multiput(10,0)(10,0){6}{\circle{3}}
\multiput(11.5,0)(10,0){2}{\line(1,0){7}}
\multiput(31.5,0)(1,0){7}{\circle*{.2}}
\put(40,0){\circle*{3}}
\multiput(41.5,0)(1,0){7}{\circle*{.2}}
\put(51.5,0){\line(1,0){7}}
\put(60,1.5){\usebox{\pole}}
\put(12,-2){\makebox(0,0)[t]{{\protect\scr 1}}}
\put(22,-2){\makebox(0,0)[t]{{\protect\scr 2}}}
\put(32,-2){\makebox(0,0)[t]{{\protect\scr 3}}}
\put(42,-2){\makebox(0,0)[t]{{\protect\scr {\em k}}}}
\put(52,-2){\makebox(0,0)[t]{{\protect\scr {\em n}--2}}}
\put(62,-2){\makebox(0,0)[t]{{\protect\scr {\em n}--1}}}
\end{picture}}
\newsavebox{\SAnTBA}
\sbox{\SAnTBA}{\begin{picture}(52,5)(0,-3.5)
\multiput(10,0)(10,0){6}{\circle{3}}
\multiput(11.5,0)(10,0){2}{\line(1,0){7}}
\multiput(31.5,0)(1,0){8}{\circle*{.2}}
\multiput(41.5,0)(1,0){8}{\circle*{.2}}
\put(51.5,0){\line(1,0){7}}
\put(12,-2){\makebox(0,0)[t]{{\protect\scr 1}}}
\put(22,-2){\makebox(0,0)[t]{{\protect\scr 2}}}
\put(32,-2){\makebox(0,0)[t]{{\protect\scr 3}}}
\put(42,-2){\makebox(0,0)[t]{{\protect\scr {\em k}}}}
\put(40,0){\circle*{3}}
\put(52,-2){\makebox(0,0)[t]{{\protect\scr {\em n}--3}}}
\put(62,-2){\makebox(0,0)[t]{{\protect\scr {\em n}--2}}}
\end{picture}}
\newsavebox{\BBBTBA}
\sbox{\BBBTBA}{\begin{picture}(52,5)(0,-3.5)
\multiput(10,0)(10,0){5}{\circle{3}}
\multiput(10,0)(10,0){1}{ \line(1,0){7} }
\multiput(21.5,0)(1,0){8}{\circle*{.2}}
\multiput(31.5,0)(1,0){8}{\circle*{.2}}
\put(41.5,0.5){\line(1,0){7}}
\put(41.5,-0.5){\line(1,0){7}}
\put(50,1.5){\line(0,1){7}}
\put(50,-1.5){\line(0,-1){7}}
\put(50,10){\circle{3}}
\put(50,-10){\circle{3}}
\end{picture}}
\hfill Torino preprint DFTT-20/94\\
\vskip 1cm
\begin{center}
{\bf
The sine-Gordon model as   $\SO(n)_{1} \times \SO(n)_{1} \over \SO(n)_{2}$
 perturbed coset theory
and generalizations
}\\
\vskip 1.8cm
{\large  R.\ Tateo }
     \footnote{E-mail:  tateo@to.infn.it}\\
\vskip .8cm
     Dip. di Fisica Teorica, Universit\`a di Torino\\
                and\\
     INFN - Sez. di Torino \\
\vskip .6cm
     Via P.Giuria 1, I-10125 Torino, Italy\\
\end{center}
\vskip 1cm
\noindent
%
Abstract:
The ground state of the
$\SO(2n)_{1} \times \SO(2n)_{1} \over \SO(2n)_{2}$ coset theories,
perturbed by the
$\phi^{id,id}_{adj}$ operator and those
of the   sine-Gordon theory, for special values of the coupling
constant in the attracting regime, is the same.
In the  first part of this paper we extend these results to the $\SO(2n-1)$
cases.
In the second part, we analyze the Algebraic Bethe Ansatz
procedure for special
points in the repulsive region. We find a one-to-one
``duality'' correspondence between these theories and those studied in the
first
part of the paper. We use the gluing procedure at the massive node proposed
by Fendley and Intriligator
in order to obtain the  TBA systems for the generalized parafermionic
supersymmetric sine-Gordon model.
In the third part we  propose  the TBA equations for
the whole class of perturbed coset models $G_k \times G_l \over G_{k+l}$
with the operator  $\phi^{id,id}_{adj}$
and $G$ a  non-simply-laced group generated by one of the
$\G_2,\F_4,\B_n,\C_n$ algebras.
\newpage
\section{Introduction}
Since the Zamolodchikov's seminal paper~\cite{zamzam}, the
two-dimensional integrable
field
theories ($IQFT_2$) have been the object of intense analysis.
In this framework, the role
played by the conformal field theories is central. The conformal field
theories are a subclass  of two-dimensional
integrable quantum field theories corresponding to the   ultraviolet (UV)
or infrared (IR) fixed point limit of the  renormalization group
trajectory  of any quantum field theories.
An $IQFT_2$ possesses an infinite set of conserved charges, and the
constraints due to
these conservation laws on the scattering process are, in any case, very
strong:  there is no particle production.
Absence of particle production implies that the S-matrix is factorizable, so
the S-matrix for any non trivial scattering is given by the product of all
possible two particles scattering amplitudes. Further, the two particles
S-matrix
elements must obey  the unitary and crossing symmetry constraints as well as
 the factorization  equation  i.e. the Yang-Baxter-Faddeev-Zamolodchikov
equation.
One of the more studied theories is the sine-Gordon (SG) model. The SG
theory  is defined by the Lagrangian
\eq
L= { 1 \over 2 } \de_{\mu} \phi \de^{\mu} \phi + { g \sqrt{ 4 \pi} \over \beta}
\cos\lf({\beta \over \sqrt{4 \pi}} \phi \ri) \pu
\en
This model has an infinite number of conservation laws. Its dynamics is simple,
but non trivial and  the factorizable S-matrix was obtained in
ref.~\cite{zamzam}.
The ultraviolet limit of the theory corresponds to  a single
free massless  boson. A massless non interacting boson $\phi$
is in general compactified  on a
circle of radius R ($ \phi \equiv \phi+ 2 \pi R$). As a consequence,
it  is necessary to include for R $\ne \infty$
 the instanton sectors and  we have to deal with a non trivial
one-parameter theory that defines
\footnote{ A  more careful analysis shows~\cite{gin1} two critical
lines which are one the orbifold of the other.}
the so-called Ginsparg line~\cite{gin1}.
Varying the parameter $R$, we reach different points on the Ginsparg line, and
 there are particular points  with a special symmetry such as
supersymmetry, discrete symmetry and Kac-Moody symmetry~\cite{gko}.
The SG model can be seen as a integrable deformation
of this  theory.  In the limit
$m^2={g \beta \over \sqrt{4\pi}} \sim 0$
we have the identification
\eq
L_{SG}(\beta) \equiv  L_{Boson}(2R= \beta/4 \pi) +\lambda \int dx^2 \Phi(x)
\virg
\en
the operator $\Phi$ has  conformal dimensions
$h(\Phi)=\bar{h}(\Phi)= {\beta^2 \over 8 \pi}$ and $|\lambda| \propto
m^{2\lf( 1- {\beta^2 \over 8 \pi} \ri)}$.
For special values of $\beta$, some unbroken symmetries of the
conformal theory  show themselves  in the perturbed theory.
This is  the case, for example,  of the supersymmetry studied in
refs.~\cite{fs,fi}.
One~\cite{mus3} of the  most effective methods found so far for  recovering
the UV behaviour of a theory, whose
factorizable S-matrix is given, is the so-called Thermodynamic Bethe Ansatz
(TBA). That the thermodynamics of a scattering theory can be reconstructed
completely from its S-matrix was known since the end of the sixties~\cite{bdm},
and the use of Bethe Ansatz technique to implement this program for integrable
theories was written in a seminal paper by Yang and Yang~\cite{yy} for a
non-relativistic scattering problem. More recently in ref.~\cite{al1}
Al.Zamolodchikov  has
proposed this method to investigate factorized scattering theories
corresponding
to $IQFT_2$.
The TBA can be presented as a set of coupled non-linear integral equations
describing the evolution of the Casimir energy of the theory on a cylinder
along
the RG flow. In spite of their apparent complexity, they are often
numerically integrable without using very heavy computer resources, for each
point on the RG flow, and show the peculiar property to be analytically
solvable in the UV and IR limits, thanks to transformations leading to
sum-rules of the Roger dilogarithm function.
In the present paper, we study
the SG theory  in some  special points. {}From the point of view of the
free boson theory, these are the
points with underlying coset $\SO(n)_1 \times \SO(n)_1 \over \SO(n)_2$
symmetries \footnote{ By considering the ground state TBA we can not
distinguish a model from its orbifold. This could be done by considering
excited states TBA, instead.}.
For this series of points we study, in the TBA framework, the property of the
perturbed theory and its relations with the $\SO(n)$ group.
This study permits us to propose the TBA equations  for all the coset models
$G_k \times G_l\over G_{k+l}$ perturbed by their $\Phi^{id,id}_{adj}$
relevant
operators  and  $G$ a group generated by one of the  $\G_2,\F_4,\C_n,\B_n$
simple Lie Algebras.
\section{The  TBA equations for the sine-Gordon at
${\beta^2\over 8 \pi}={2 \over 2n+1}$}
The SG S-matrix is the minimal~\cite{zamzam}
 $O(2)$-symmetric  solution  of the unitary, crossing and
factorization equations. The two-particle S-matrix
amplitudes for the scattering of  a soliton  $A$ and the anti-soliton
$\bar{A}$ can be written as
\bea
A(\th_1)A(\th_2) &=& S(\th_1-\th_2,\xi)A(\th_2) A(\th_1)
\nn \\
\bar{A}(\th_1)\bar{A}(\th_2)&=&S(\th_1-\th_2,\xi)
\bar{A}(\th_2)\bar{A}(\th_1)  \\
A(\th_1)\bar{A}(\th_2)&=&S_T(\th_1-\th_2,\xi) \bar{A}(\th_2)A(\th_1)+
S_R(\th_1-\th_2,\xi) A(\th_2)\bar{A}(\th_1) \pu\nn
\eea
The soliton-antisoliton S-matrix $\hat{S}$  in a matrix form is
\[
\hat{S}\lf(\th, \xi \ri)= S_0\lf(\th, \xi \ri) \hat{R}\lf(\th, \xi \ri)=
\lf(
\begin{array}{llll}
   S(\th,\xi)     &                &                &                \\
                  &  S_R(\th,\xi)  &  S_T(\th,\xi)  &                \\
                  &  S_T(\th,\xi)  &  S_R(\th,\xi)  &                \\
                  &                &                & S(\th,\xi)     \\
\end{array} \ri)
\]
\[
\hat{R}(\th,\xi)=
\lf(
\begin{array}{llll}
   {\sinh\lf({\pi \over \xi} (\th- \imath \pi) \ri)} & & & \\
                  & -{\sinh \lf({\imath \pi^2 \over \xi} \ri) } &
-{\sinh\lf( {\pi \over \xi} \th \ri)} & \\&
  -{\sinh\lf( {\pi \over \xi} \th \ri)}
 & -{\sinh \lf({\imath \pi^2 \over \xi} \ri) } &  \\
 & & &  { \sinh\lf( {\pi \over \xi} (\th- \imath \pi) \ri)} \\
\end{array}
\ri)\]
\eq
S_0(\th,\xi)= {1 \over  \sinh\lf({\pi \over \xi} (\th- \imath \pi)\ri)}
\exp \lf[ - \imath  \int_0^{+\infty}
{dk \over k} { \sin(k\th) \sinh\lf({{\pi - \xi} \over 2}k\ri)
\over \cosh\lf({\pi k \over 2}\ri) \sinh\lf({\xi k \over 2} \ri)} \ri]
\label{S0}
\en
and
\eq
\xi= {{ \beta^2 \over 8} \over {1 - { \beta^2 \over 8 \pi }}}
\virg
\en
the poles in $S_0$
for $\th=\imath \pi - \imath j \xi$~  $j=1,2, \dots < \pi/\xi$
are interpreted as $A~\bar{A}$   bound state.
They are the  well-known  breathers in the  SG  model. The breathers
masses  are
\eq
M_j=2 M_n \sin\lf({j \xi \over 2} \ri)\sps \sps
j=1,2,\dots  {< {\pi \over \xi} }
\pu
\label{mass}
\en
In eq.(\ref{mass})  $M_n$ is the soliton mass.
The two particle  S-matrix for the scattering  breather-soliton and
breather-breather can be obtained using the bootstrap equations.
We have
\bea
S^{i,A}_{i,A}(\th,\xi) &=&
{ \sinh \th + \imath \cos\lf({ i \xi \over 2}\ri) \over
 \sinh \th - \imath \cos\lf({ i \xi \over 2}\ri) }
\prod_{l=1}^{i-1} { \sin^2\lf( { i-2l \over 4} \xi -
{\pi \over 4} + \imath { \th \over 2} \ri)     \over
 \sin^2\lf( { i-2l \over 4} \xi - {\pi \over 4} -
\imath { \th \over 2} \ri)}  \label{sai}\\
S^{i,k}_{i,k} (\th, \xi) &=&
{ \sinh \th + \imath \sin\lf({ i+k  \over 2}\xi\ri) \over
  \sinh \th - \imath \sin\lf( { i+k  \over 2}\xi\ri) }
{ \sinh \th + \imath \sin\lf({ i-k  \over 2}\xi\ri) \over
  \sinh \th - \imath \sin\lf( { i-k  \over 2}\xi\ri) }
\times \nn \\
 &\times&
  \prod_{l=1}^{min(i,k)-1}
{ \sin^2\lf( { k-i-2l \over 4} \xi  + \imath { \th \over 2} \ri)
 \cos^2\lf( { k+i-2l \over 4} \xi  + \imath { \th \over 2} \ri)     \over
 \sin^2\lf( { k-i-2l \over 4} \xi  + \imath { \th \over 2} \ri)
    \cos^2\lf( { k+i-2l \over 4} \xi  + \imath { \th \over 2} \ri)  }
\pu
\label{sij}
\eea
We are interested in  the SG theory  at  $\xi= { 2 \pi \over m}$ and
$m \in Z^+$.
In
$m=2(n-1)$   the reflection
coefficient
$S_R$ vanishes and the theory  is  purely elastic. A deep analysis
can be found in refs.~\cite{km1,km2,dr,dr1}.
The ground state in these points
is equivalent to that of  the $\SO(2n)_1 \times \SO(2n)_1 \over \SO(2n)_2$
perturbed  coset
theories. The perturbing operator  is the field
$\Phi^{id,id}_{adj}$, and its conformal dimensions are
\eq
h(\Phi^{id,id}_{adj})=\bar{h}(\Phi^{id,id}_{adj})=
1 - {\tilde{h} \over  \tilde{h}+2} =
{1 \over n}
\pu
\label{peso}
\en
In eq. (\ref{peso}) $\tilde{h}$ is the  dual $\SO(2n)$
Coxeter number and $\tilde{h}$ is $n-2$ for  $\SO(n)$.
We are interested in  completing  the study of the  theories
\eq
{\SO(m)_1 \times \SO(m)_1 \over \SO(m)_2}+ \Phi^{id,id}_{adj}
\sps \sps \sps h(\Phi^{id,id}_{adj})=\bar{h}(\Phi^{id,id}_{adj})=
 {2  \over m}
\label{SO}
\en
with   $m=2n+1$ that is the orthogonal groups generated by the $\B_n$ Lie
algebra. We based our analysis on the ground state equivalence
of these theories
and the  SG  for $\xi={2 \pi \over 2n-1}$ and
we  applied the Algebraic Bethe Ansatz (ABA) procedure
in order to extract information about
the vacuum energy of the field theory at finite temperature
\footnote{After the completing of this article  H.Itoyama informed us that
a variant of  the sine-Gordon theory at these points emerges
in connection with
the higher spin $N=2$ supersymmetry~\cite{it}.
In that contest also  H.Itoyama and T.Oota used
the results obtained in~\cite{fi} for the
six-vertex model with free fermion conditions in order to obtain
a TBA system.}.
Let us write the amplitudes
(\ref{S0},\ref{sai},\ref{sij}) at
$\xi={2 \pi \over 2n-1}= { 2 \pi \over \tilde{h}}$ in a more convenient form
\bea
S_0(\th)&=&
 {(-1)^{n} \imath \over  \cosh\lf( {\tilde{h} \over 2} \th \ri)} \times
\nn \\
            & \times &         \exp \lf[ - \imath  \int_{-\infty}^{+\infty}
{dk \over k} { \sin(k\th)  \over
4\cosh^2\lf( {\pi k \over  2\tilde{h}}\ri)}
{ \lf[ {  T
\over 2 \cosh\lf( {\pi k \over \tilde{h}} \ri) - T} \ri]_{n-1,n-1}} \ri] \nn\\
S_{A,m}^{A,m}(\th) &=&
 \exp \lf[ - \imath  \int_{-\infty}^{+\infty}
{dk \over k} { \sin(k\th)  \over
2 \cosh\lf( {\pi k \over  2\tilde{h}}\ri)}
{ \lf[ {2 \cosh\lf( {\pi k \over  \tilde{h}}\ri)
\over 2 \cosh\lf( {\pi k \over \tilde{h}} \ri) - T} \ri]_{m,n-1}} \ri]
\label{smat} \\
S_{n,m}^{n,m}(\th)&=&
\exp \lf[ - \imath  \int_{-\infty}^{+\infty}
{dk \sin(k \th)\over k} { \lf[ {T
\over 2 \cosh\lf( {\pi k \over \tilde{h}} \ri) - T} \ri]_{m,n}} \ri]\pu \nn
\eea
In (\ref{smat}) $T_{ab}$ is the incidence matrix
of the tadpole  diagram $\T_n$
(see {\em Figure~\ref{fig11}}). \\

\begin{figure}[htbp]
\begin{center}
\begin{picture}(130,70)(0,0)
\put(-10,45) {\usebox{\Tn}}
\put(55,43){{\stl $M_i=2 M \sin\lf( {\pi i \over (2n-1)}
\ri)$,\sps $i=1,\ldots,n-1$}}
\put(-20,20){\parbox{130mm}{\caption{\label{fig11}\protect {\sts
The $\T_{n-1}$ diagram  is associated  with the bosonic  sub-sector
of the theory.
This is the only sector that survives after  the quantum  reduction.
The reduced theories are the minimal non--unitary
 $M_{2,2n+1}$ perturbed via the fields $\Phi_{13}$.}}}}
\end{picture}
\end{center}
\end{figure}
\noindent
The first equation
was obtained  in refs.~\cite{al2,rtv}.
We obtain the second and the third from the bootstrap properties
\footnote{
\eq
\Th(x)=\lim_{\epsilon\to 0}\lf[\frac{1}{2}+\frac{1}{\pi}\arctan {x \over
\epsilon} \ri] =
\lf\{ \begin{array}{lll}
0 & {\rm if} & x<0 \\
\frac{1}{2} & {\rm if} & x=0 \\
1 & {\rm if} & x>0
\end{array} \ri.
\en}
\eq
S_{i,{n-1}}^{i,{n-1}}(\th) e^{\imath 2\pi\delta_{i,n-1} \Th(\th)}=
S_{A,i}^{A,i}\lf(\th + { \imath \pi \over 2 \tilde{h}} \ri)
S_{A,i}^{A,i}\lf(\th - { \imath \pi \over 2 \tilde{h}} \ri)
\en
\eq
\ba{c}
\ds{ S_0 \lf(\th + {\imath \pi \over 2 \tilde{h}} \ri)
S_0\lf(\th - {\imath \pi \over 2\tilde{h}} \ri) = S_{A,{n-1}}^{A,{n-1}}(\th)
\times} \acc
 \ds{ \sps \sps \times { 1
\over
\sinh\lf( { \tilde{h} \th \over 2} - { 3 \imath \pi \tilde{h} \over 4} +
 { \imath \pi \over 2} \ri)
\sinh\lf( { \tilde{h} \th \over 2} + { \imath \pi \tilde{h} \over 4} +
{ \imath \pi \over 2} \ri) }}\acc
\ea
\en
via a Fourier transformation.
The quantization equations for the system with $N$ solitons of mass $M_n$ and
$N_b$ breathers of mass $M_i$, $i=1,\dots,n-1$ on a circle of circumference
L  with periodic boundary conditions are
\bea
e^{\imath L M_n \sinh \th} \Lambda(\th|\th_1,\dots,\th_N)
\prod_{k=1}^{N_b} S^{A,k}_{A,k}(\th-\th_k)
&=& 1 \nn\\
e^{\imath L M_i \sinh \th} \prod_{j:j \ne i}^{N_b} S_{i,j}^{i,j}
(\th-\th_j) \prod_{k=1}^{N} S_{A,i}^{A,i}(\th-\th_k)&=&1
\label{sono}
\eea
In (\ref{sono}) $\Lambda(\th|\th_1,\dots,\th_N)$ is the eigenvalue
of the trace $T$ of the transfer matrix
\bea
T(\th | \th_1,\dots,\th_N)^{\{d_j\}}_{\{c_j\}} &=&
\sum_{k_i} \hat{R}^{d_1,k_1}_{k_N,c_1}(\th-\th_1)
 \times \nn  \\ &\times&
\hat{R}^{d_2,k_2}_{k_1,c_2}(\th-\th_2) \dots
\hat{R}^{d_N,k_N}_{k_{N-1},c_N}(\th-\th_N)
\eea
and
\[
\Lambda(\th | \th_1,..,\th_N)=
 \lf( \prod_{i=1}^{N} \sinh\lf( {\tilde{h} \over 2} (\th_i-\th) \ri) +
(-1)^{r} \prod_{i=1}^{N} {\sinh\lf( {\tilde{h} \over 2} (\th-\th_i -\imath \pi
) \ri)} \ri) \times
\]
\eq
\times \prod_{j=1}^{r} { \sinh\lf( {\tilde{h} \over 2} (\th-y_j- \imath \pi)
\ri)
 \over {\sinh\lf( {\tilde{h} \over 2}(y_j-\th )\ri)}}
\label{La}
\en
The $y_j$  must satisfy the equation~\cite{fi}
\eq
\prod_{i=1}^{N} { \sinh\lf( { \tilde{h} \over 2}(\th_i- y_j)\ri) \over
\sinh\lf( { \tilde{h} \over 2}(y_j-\th_i-\imath \pi) \ri)} = (-1)^{r+1}
\pu
\label{sinsin}
\en
The solutions of eq.~(\ref{sinsin}) are of the form
$y'=\Re e(y')+ \imath \pi/2$, $\bar{y'}=\Re e(\bar{y'})- \imath \pi/2$,
$y'=\tilde{h}y$ , $\bar{y}'=\tilde{h}\bar{y}$.
Defining in the limit $N \fr \infty$ and  $L \fr \infty$
the densities \footnote{
To be more precise, in passing from the variable $y'$ to the variable
$y$ we should
introduce an additional index labelling the pseudoparticles. However,
it is easy
to see that the  vacuum state can be correctly described introducing  only
two psedoparticle species. The resulting TBA system is in a more
universal form.}
$\rho_{(n,1)}( \Re e (y_i))$ and $ \rho_{(n,3)}(\Re e (\bar{y}_i))$ we find
\eq
\rho_{(n,i)}(\th) ={1 \over 2\pi}\int_{-\infty}^{+\infty} dz { \tilde{h}~
\rho_{(n,
2)}^{+}
(z)
\over \cosh(\tilde{h} (\th -z))}
= (\Phi * \rho_{(n,2)}^{+})(\th)
\virg
\label{den}
\en
\eq
\Phi(\th)= \int_{-\infty}^{+\infty}  dk {\cos(k \th) \over 2 \cosh \lf({ \pi k
\over 2 \tilde{h} }\ri)}
\en
in eq.(\ref{den}) $i=1,3$ and
$\rho^{\pm}_{(n,j)}$ is the density of roots (holes) of
 particles ($j=2$)  or pseudoparticles ($j=1,3$);
 $\rho_{(n,j)}=\rho^{+}_{(n,j)}+\rho^{-}_{(n,j)}$
is the  density of states (roots and holes).
{}From eqs.~(\ref{sono}) we find
\bea
\rho_{(n,2)} (\th) &=& {M_n \over 2 \pi}  \cosh \th + (\Phi_A *
 \rho_{(n,2)}^{+})(\th) + \nn \\
                   &+& {1 \over 2} \sum_{i=1,3} (\Phi *(\rho^+_{(n,i)}-
\rho^-_{(n,i)}))(\th) + \sum_{k=1}^{n-1} (\Phi_k * \rho_k^+)(\th) \label{tb} \\
\rho_j(\th) &=& {M_j \over 2 \pi} \cosh \th + \sum_{m=1}^{n-1} (\Phi_{j,m}*
\rho_m^+)(\th) +(\Phi_j * \rho^{+}_{(n,2)})(\th)
\nn
\eea
with
\eq
\ba{c}
\ds{\Phi_{j,m}(\th) = -\imath { d \over d \th} \log S_{j,m}^{j,m}(\th)
\sps \sps \Phi_{A}(\th) =  -\imath { d \over d \th} \log S_0(\th) } \acc
\ds{\Phi_{m}(\th)= -\imath { d \over d \th} \log S_{A,m}^{A,m}(\th) }
\pu
\ea
\en
In (\ref{tb})  $\rho^{+}_j(\th)$ and $\rho_j(\th)$ are
the density of roots and  states for the   particles of species $j$.
Putting
\bea
{\rho^{+}_{(n,i)}(\th) \over \rho_{(n,i)}(\th)} &=& {1 \over 1+e^{\ep_{(n,i)}
(\th)}} \sps \sps
{\rho^{+}_{j}(\th) \over \rho_{j}(\th)} =  {1 \over 1+e^{\ep_{j}(\th)}}
\eea
we obtain, after extremizing the free energy at temperature $T=1/R$,
\bea
 R M_n \cosh \th &=& \ep_{(n,2)}(\th) +  (\tilde{\Phi}_A *
 \log(1+e^{-\ep_{(n,2)}}))(\th) + \nn \\
&+& \sum_{i=1,2}(\tilde{\Phi} *
 \log(1+e^{-\ep_{(n,i)}}))(\th)
+\sum_{k=1}^{n-1} (\Phi_k *\log(1+e^{-\ep_{k}}) )(\th)\nn \\
 R M_j \cosh \th &=&\ep_j(\th) + (\Phi_j * \log(1+e^{-\ep_{(n,2)}}))
(\th)+ \nn \\
 &+&   \sum_{k}^{n-1} (\Phi_{j,k}* \log(1+e^{-\ep_k}))(\th)\sps j=1,\dots,n-1
 \label{hhh}\\
 0 &=& \ep_{(n,i)}(\th) + (\Phi* \log(1+e^{-\ep_{(n,2)}}))(\th)\sps i=1,3 \nn
\eea
with
\eq
\tilde{\Phi}_A(\th)=  - \int_{-\infty}^{+\infty}
{dk} { \cos(k\th)  \over
4 \cosh^2\lf( {\pi k \over  2\tilde{h}}\ri)}
{ \lf[ { 2 \cosh\lf( {\pi k \over \tilde{h}} \ri)
\over 2 \cosh\lf( {\pi k \over \tilde{h}} \ri) - T} \ri]_{n-1,n-1}}
\pu
\label{nuclei}
\en
The UV central charge is obtained using standard methods via the
Roger~\cite{lewin}
\footnote{
\eq
L(x)=-{1 \over2 } \int_{0}^{x} dy \lf[{\log(y) \over 1-y} +{\log(1-y)
 \over y}\ri]
\label{c26}
\en
}
dilogarithm function and its sum-rules. Using the stationary  solutions
$\ep_0$ of eqs.
(\ref{hhh}) at  $R=0$ and  putting $y_a=e^{\ep_0^{a}}$, we have
\eq
c_{uv} = \frac{6}{\pi^2} \lf[ \sum_{a~\in~all~nodes}
L\lf(\frac{1}{1+y_a}\ri)
- 2L\lf(\frac{1}{2}\ri) \ri]
\en
and
\eq
\ba{c}
\ds{ y_i = i(i+2) \sps \sps i = 1,\dots,n-1 } \acc
\ds{ y_{(n,2)}= { n^2 \over 2n+1} \sps \sps y_{(n,1)}= y_{(n,3)}= { n \over
n+1}
\pu}
\ea
\en
Making use of some property pointed out  in ref.~\cite{lewin}
\load{\footnotesize}{\sf} \footnote{ $\sum_{k=2}^{n}
L\lf( {1 \over k^2} \ri)+ 2L \lf( {1 \over n+1} \ri) ={ \pi^2 \over 6} $ and
$ L(x^2)=2L(x) -2 L \lf( { x \over 1+x} \ri)$ \pu}
we find  $c_{uv}=1$ as expected. \\

\begin{figure}[htbp]
\begin{center}
\begin{picture}(130,70)(0,0)
\put(-10,30) {\usebox{\Bn}}
\put(59,40){{\stl $M_i = 2 M \sin\lf( {\pi i \over (2n-1)}
\ri)$,\sps $i=1,\ldots,n-1$}}
\put(63,30){{\stl $M_n = M $ }}
\put(-20,10){\parbox{130mm}{\caption{\label{fig1} \protect {\sts
$\B_n$ diagrams. }}}}
\end{picture}
\end{center}
\end{figure}

\vspace{0.3cm}

\begin{figure}[htbp]
\begin{center}
\begin{picture}(130,50)(0,-20)
\put(35,5) {\usebox{\BnTBA}}
\put(-20,-15){\parbox{130mm}{\caption{ \label{fig2}\protect {\stl
The  graph associate to the  $\B_n$  TBA equation. The black nodes
correspond to
particles, the white nodes  to pseudoparticles.}}}}
\end{picture}
\end{center}
\end{figure}
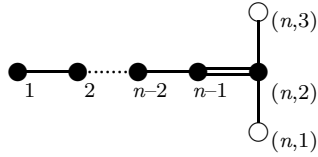

\noindent
In order to find in a simple way  the conformal dimension of the perturbing
field it is convenient to move
from the  TBA equation to the  Y-system associate equation~\cite{al2}.
Using the following   mass spectrum  properties
\load{\footnotesize}{\sf} \footnote{
This property generalizes the well-known  property which states that the mass
spectrum in the $\A\D\E$ scattering theories are the components of the
Perron Frobenius eigenvector  of the incidence matrix of the associate
Dynkin
diagram.}
\bea
 2 \cos\lf( { \pi \over  \tilde{h}} \ri) M_i &=& M_{i-1}+ M_{i+1}\sps \sps i
 \le n-2 \nn \\
 2 \cos\lf( {\pi \over \tilde{h}} \ri) M_{n-1} &=&
M_{n-2} + 2 \cos\lf( {\pi \over 2
\tilde{h}} \ri) M_n \label{www} \\
 2 \cos\lf( {\pi \over 2 \tilde{h}} \ri) M_n &=& M_{n-1} \nn
\eea
we find
\[
 Y_i \lf( \th + \imath { \pi \over \tilde{h}} \ri)
Y_i \lf( \th - \imath { \pi \over \tilde{h}}\ri) =
(1+Y_{i-1}( \th ))(1+Y_{i+1}(\th)) \sps i=1,\dots n-2 \nn
\]
\eq
\ba{c}
\ds{ Y_{n-1} \lf( \th + \imath { \pi \over \tilde{h}} \ri)
Y_{n-1} \lf( \th - \imath { \pi \over \tilde{h}}\ri) =
(1+Y_{(n,1)} (\th))(1+Y_{(n,3)} (\th))(1+Y_{n-2}( \th )) \times }\acc
\ds{ \times   \lf(1+Y_{(n,2)} \lf(\th + \imath
{ \pi \over 2 \tilde{h}}
\ri)\ri)
\lf(1+Y_{(n,2)} \lf(\th - \imath { \pi \over 2\tilde{h}}\ri) \ri)
}
\label{yyy}
\ea
\en
\bea
Y_{(n,2)} \lf( \th + \imath { \pi \over 2\tilde{h}} \ri)
Y_{(n,2)} \lf( \th - \imath { \pi \over 2\tilde{h}}\ri)&=&
 \lf( 1 + {1 \over Y_{(n,1)}(\th)} \ri)^{-1}
\lf( 1 + {1 \over Y_{(n,3)}(\th)} \ri)^{-1} \times \nn \\
&\times& (1+Y_{n-1}(\th)) \nn
\eea
\[
Y_{(n,i)} \lf( \th + \imath { \pi \over 2\tilde{h}} \ri)
Y_{(n,i)} \lf( \th - \imath { \pi \over 2\tilde{h}}\ri) =
\lf( 1 + {1 \over Y_{(n,2)}(\th)} \ri)^{-1} \sps \sps i=1,3
\pu
\]
In eq. (\ref{yyy}) $Y(\th)=e^{\ep(\th)}$.
It is possible to recursively demonstrate for the first few cases,
and verify numerically
 in the others,   that the Y-system (\ref{yyy}) defines
functions  $Y(\th)$ with periodicity
\eq
Y_a(\th)=Y_a(\th+ \imath \pi P) \sps \sps
P= { \tilde{h}+2  \over \tilde{h}}
\sps \sps \forall a
\label{period}
\en
$P$ is in relation with the conformal dimension of the perturbing field
$\Phi$ via the formula
\eq
h(\Phi)=\bar{h}(\Phi)= {\beta^2 \over 8 \pi} =1 -  { 1 \over P}
\pu
\en
This confirms our identification and the correct procedure
in   Bethe Ansatz's framework.
Before concluding this section, we would like to give a graphic interpretation
for the TBA systems  obtained.
Differently from the diagrammatic interpretation in the $\A\D\E$ cases
{}~\cite{rsos,altim,alcoset,wtba,fatal,fi,zamzam1,al4,rtv},
  this graphic picture is not completely  rigorous\load{\footnotesize}
{\sf}
\footnote{
However, it is possible to define some simple
path-like rule in order to move from the graph to the Y-system, and, after a
Fourier transformation involving the mass spectrum properties
(\ref{www}), to the associate TBA
system.}, it  permits only a
schematization of the TBA equations, but  it helps to give a
more intuitive meaning to
the topic we
are dealing with.
Because the  presence of pseudoparticles in the theory
it is convenient to introduce a graph
with a number of nodes equal to the number of particles species ( black  nodes)
plus the number of pseudoparticles ones ( white nodes ).
We define a TBA graph
adding two magnonic nodes to the Dynkin diagram in correspondence to the
soliton
node $n$. This node is  the
lowest root
node in the Dynkin diagram.  The result is represented in
{\em Figure~\ref{fig2}}.
\acc
{\em Comments:}
In this section we have used the sine-Gordon S-matrix  at
$ \xi= {2 \pi \over m}$  to study the
ground state  of the theory (\ref{SO}). The
$\B_n$ S-matrix  can be
obtained from the SG S-matrix through lattice model like orbifold~\cite{gin}.
Once  the S-matrix is obtained one can hope to dress it  by a
Z-factor~\cite{mus,mus1} containing only zeros
 in the physical strip  to obtain the
S-matrix for the associate Toda field theory. In this
way  the S-matrix proposed in ref.~\cite{des} for the Toda $B_n$
theory should be
obtained. This work is in progress~\cite{tat}.\hfill
\resection{ On the  TBA equations for the  supersymmetric sine-Gordon models}
 The study of the SG and fractional supersymmetric sine-Gordon models
(FSSG)~\cite{dbal} in the repulsive regime is very interesting.
In this section  we shall study the FFSG
for the values
\eq
{ \beta^2 \over 8 \pi} = {1 \over k} - {2 \over m}
\label{VV}
\en
$ k=1,2,\dots \le (int[m/2]-1)$ and $ m=1,2,\dots$ .
These theories  are related
to those
studied in the preceding section by a sort of TBA duality that permits the
passing
from the TBA equations of a  model in
the repulsive regime to a "dual'' model in the attractive regime.
The FSSG S-matrix can
be obtained by the SG S-matrix using the  gluing procedure at the massive node
proposed in ref.~\cite{fi}.
The  FSSG model  is an interacting theory of a boson
and a generating parafermion $\Z_k$.
The case  $k=1$ corresponds to  the  SG model,
the case $k=2$ is a supersymmetric $N=1$ theory, in this case
 the fundamental
interacting fields in the Lagrangian
are  a Majorana fermion and a  boson.
The interacting vertex in the FSSG model is
\eq
V(\phi,\Psi_1)= {g \sqrt{4 \pi}\over \beta}  \Psi_1(z)\bar{\Psi}_1(\bar{z})
e^{-\imath {\beta  \over \sqrt{4\pi}} \phi(z,\bar{z})}
\label{ver} + c.c
\pu
\en
In (\ref{ver}) the field
$\Psi_1(z)$ is the generating parafermion  with conformal dimension
$h(\Psi_1)={ k-1 \over k}$, and the  conformal weights of
the vertex  $e^{-{\imath \beta
\over \sqrt{4 \pi}} \phi}$ are $\lf( { \beta^2 \over 8
\pi},{ \beta^2 \over 8 \pi}\ri)$.
The theory is integrable
and has both  local and non-local conserved currents.
The UV limit is a theory of a boson and a parafermion at the compactified
radius
$R= {\sqrt{4 \pi} \over \beta}$. The renormalized coupling constant is
\eq
\xi(\beta)= { k {\beta^2/ 8 } \over {1/k - \beta^2/8 \pi}}
\pu
\en
The S-matrix is the tensorial product of the SG S-matrix $S_{SG}$ and
the restricted SG S-matrix
$S^K_{RSG}$ i.e. the $\Z_k$ parafermionic theory obtained by
  the  massive $\Phi_{13}$
perturbation of the minimal model $\M_{k-1}$
\eq
S^K_{FSSG}(\th)=S^K_{RSG}(\th) \otimes S_{SG}(x=e^{\th \pi/\xi},q=-e^{-\imath
/ \xi})
\label{ssss}
\en
The  spectrum  consists in  a  soliton and antisoliton
pair with mass $M_A$ and neutral bound states with masses
\eq
M_j=2M_A \sin\lf( {j  \xi \over 2} \ri), \sps j=1,2,\dots, < {\pi \over
\xi}
\pu
\en
We are interested in  this theory for the values given in (\ref{VV}).
For  $k=1$ and $m=2n$ the system has been  studied in ref.~\cite{fatal},
these points correspond to the SG theory. The  case $k=n-2$ and
$m=2n$  was  studied in
{}~\cite{fi} and corresponds to   particular  points
with supersymmetry  $N=2$.   The  TBA  for $k=2,3,\dots, n-3$ and $m=2n$
was proposed in ref.~\cite{rtv}. All these  TBA  theories  are of the magnonic
type
and encoded on  a  $\D_n$ Dynkin diagram  (see
{\em Figure~\ref{fig10}}). This common structure
derives  from  equation
(\ref{ssss}) and in the TBA framework corresponds
to the fusion of two graphs at the massive node
(compare {\em Figure~\ref{fig10}}  and {\em Figure~\ref{fig13}}
for $k=1$ with {\em Figure~\ref{fig10}} ).
In order to complete
this series
one has to study  the points with $k=n-1$,
in  SG  this is the
free Dirac point and the corresponding TBA diagram consists in two
non-connected
nodes. Because the mass  of the fundamental boson at fixed $\beta$
is the  only   free parameter in SG,
the two nodes are both massive
and with identical mass. Because  the TBA
for the vacuum energy
can not distinguish a model from its orbifold  this TBA  describes
also the
$Ising^2$ system i.e. two free  massive
Majorana fermions. We
identify the two nodes in the two independent self-conjugate
fermions \footnote{This means that,  in the  orbifolded system,  we can have
two different temperature
perturbations  and the TBA equations  consist in two independent nodes
with
different mass terms}. It seems possible to generalize this to  all the
TBA in {\em Figure~\ref{fig10}}  with
$k=n-1$. This corresponds to a graph with all magnonic nodes,
except the nodes $n-1$ and $n$ with masses
 $m_{n-1}=m_{n}$
\footnote{ Also in this case it seems  possible to have  $m_{n-1} \ne m_{n}$.
This should  correspond
to a two  parameter perturbation of the orbifolded system at the
generalized free Dirac point.}.   The renormalized coupling constant is
$\xi= \pi$
which is  independent of $n$.
We deal now  with the
case  $m=2n+1$. We do not report the complete calculus which is very similar
 to
those reported in the first part of this work, but
let us summarize the results obtained from  our ABA
analysis of the SG model at these points.
It turns out that the TBA equations  are
described by  a particle with mass $M_A$
and $n+1$
pseudoparticles. A peculiarity of these TBA  is to have
an
antisymmetric integral kernel (for instance the $\Phi_{n-1,n}$ ones).
In spite of  this,
it is simple to
obtain a Y-system from the TBA equations which  is similar to
those proposed in the preceding section (see eq.(\ref{yyy}) )
with the following   modifications
\[
\tilde{h} \rightarrow 2
\]
\eq
Y_a(\th)  \rightarrow Y_a^{-1}(\th)  \sps\sps\sps a=1,\dots,n-1
\label{ca}
\en
\[
Y_{(n,i)}(\th)  \rightarrow Y_{(n,i)}(\th)  \sps\sps i=1,2,3
\pu
\]
One can obtain a symmetric TBA defining
\bea
\tilde{Y}_{(n,i)}(\th)  &=& Y_{(n,i)}^{-1}(\th) \sps \sps i=1,2,3 \nn \\
\tilde{Y}_a(\th)      &=& Y_a(\th) \sps\sps\sps a=1,\dots,n-1
\virg
\eea
so,  we  are dealing with a diagram like that in
{\em Figure~\ref{fig2}} but now of magnonic type
(see {\em Figure~\ref{fig9}} for $k=1$).
{}From the equation (\ref{ssss})
it follows that   the TBA systems in {\em Figure~\ref{fig9}}
with mass term on the node   $k$ correspond
to the generalized  FSSG model at the points (\ref{VV}) with $m=2n+1$.
For  $k=n-1$ we find, also in this case,  special properties. At these points
we have  $\xi=\pi/2$ and  there is a neutral bound state
$A \bar{A}$ with mass  $\sqrt{2}M_A$.
At this point  the SG theory ($k=1$) is
reflectionless and  the  orbifolded theory contains three interacting
 $\Z_4$ symmetric bosons.
\acc
{\em Comments:}
In this section
we have noted, in the SG theory,  a sort
of duality relation from the repulsive regime to the attractive.
The steps involved in moving from a theory to its ''dual`` are the following
\eq
\tilde{h} \Leftrightarrow  2 \sps \sps
Y(\th) \Leftrightarrow \tilde{Y}^{-1}(\th) \sps \sps
M_i \Leftrightarrow 0 \sps i=2, \dots \sps \sps
M_1 \Leftrightarrow M_A
\label{ca1}
\pu
\en
After this procedure we have
\eq
\xi \Leftrightarrow  {1 \over \xi} \sps \sps
\Delta_{per} \Leftrightarrow 1-\Delta_{per} \pu
\en
Also in the cases $m=2n$ this scheme is  valid
and  we have checked the validity of this procedure
for other  rational points. We do not have a general
method  for the ABA procedure, but in the attracting  region, where we obtain
a solution,
we observe a more complicated  magnonic structure. To understand
this structure
we can think to diagram
like that in  {\em Figure~\ref{fig2}}, but with the nodes
$(n,i)$ replaced by a graph like those
represented in {\em Figure~\ref{fig9}} or  those in  {\em Figure~
\ref{fig10}}
 for $k=1$.
On the other hand, in the repulsive regime we find a structure
identical
to that in the attracting regime via a change like (\ref{ca1}).
This  procedure, whose validity can be checked a posteriori via the
perturbation theory, can be
used to find the TBA equations for the minimal model in points in which
it is not
possible to quantum reduce the SG model.
Once  the TBA system for the reduced model at the point
\eq
{ \xi \over \pi} = {p \over q-p}
\virg
\en
it is know, the TBA system for
the dual model
\eq
{ \tilde{\xi} \over \pi} = {\pi \over \xi}= {\tilde{p} \over
\tilde{q}-\tilde{p}}
\en
can be determined without passing through the   reduced S-matrix,
provided   one determines the duality transformation completely.\hfill
\begin{figure}[htbp]
\begin{center}
\begin{picture}(130,40)(0,-20)
\put(0,5) {\usebox{\SBnTBA}}
\put(70,5) {\usebox{\SBBnTBA}}
\put(-20,-15){\parbox{132mm}{\caption{\label{fig9} \protect {\sts
A diagram of the
$\B_n(TBA)$ type is associated with the supersymmetric sine-Gordon theories
with
parafermionic supersymmetry  $\Z_k$: $k$ is the position of the massive node.
}}}}
\end{picture}
\end{center}
\end{figure}
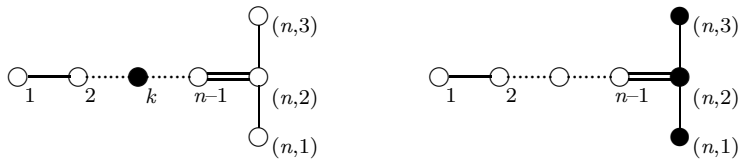
\begin{figure}[htbp]
\begin{center}
\begin{picture}(130,40)(0,-20)
\put(0,5) {\usebox{\SDnTBA}}
\put(70,5) {\usebox{\SDDnTBA}}
\put(-20,-15){\parbox{132mm}{\caption{\label{fig10}\protect {\sts
A diagram of the  $\D_n(TBA)$ type is associated with the
supersymmetric sine-Gordon theories with parafermionic
supersymmetry  $\Z_k$: $k$ is the position of the massive node. }}}}
\end{picture}
\end{center}
\end{figure}
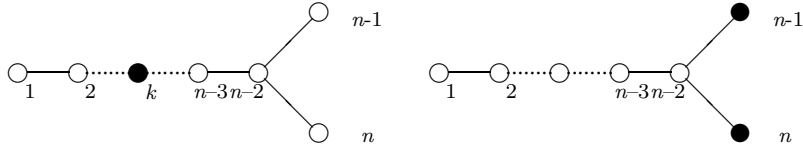
\resection{Quantum reduction: flux between  the   models}
At the points (\ref{VV}) with $m=2n$ it is possible to operate a
 quantum reduction~\cite{sm} of the
model,  this allows to obtain  the perturbed
$SU(2)_k \times SU(2)_l \over SU(2)_{k+l}$  coset theories with $k+l=n-1$.
The  TBA diagram is represented in
{\em Figure~\ref{fig13}.a}. In the case  $k=1$ we have the minimal  theories
$\M_n + \Phi_{13}$. The renormalization fluxes~\cite{fsz} between
different  points
in the  FSSG correspond, after the quantum reduction,
to fluxes\acc
\eq
{\SU(2)_k \times \SU(2)_l \over \SU(2)_{k+l}} + \Phi_{per}
\fr {\SU(2)_{l-k} \times \SU(2)_k \over \SU(2)_{l}}+\Phi_{act}
\label{f1}
\en
with
\eq
\Delta( \Phi_{per} \equiv \Phi^{id,id}_{adj})=
\lf(1-{2 \over k+l+2},1-{2 \over k+l+2}\ri) \sps\sps
\label{f2}
\en
\eq
\Delta( \Phi_{act})=\lf(1+{2 \over l-k+2},1+{2 \over l-k+2}\ri)
\pu
\label{f3}
\en
At the points $m=2n+1$ the quantum reduction
does not work. As consequence we can not
consistently define the S-matrix for the reduced model from the FSSG theory.
However, using the
approach proposed in ref.~\cite{fs}, we
obtain the $\M_{p,q}$ ground-state TBA equations
adding an
appropriate chemical potential
to the sine-Gordon TBA
at $\xi={ p \pi \over q-p}$. We generalize this  procedure to
the  FSSG model. Introducing in  our theories   a
chemical
potential in the same way of ref.~\cite{fs},
we obtain  that the three
nodes
$(n,i)$ decouple from the rest of the system and the associate TBA corresponds
 to
the $\T_{n-1}$ TBA reported in {\em Figure~\ref{fig13}.b}. We
note that this is the same
 substitution done in the attracting case (see {\em Figure~\ref{fig11}}).
So, from the substitution  $\B_n \fr \T_{n-1}$, we obtain the
TBA equations for the theories with fractional supersymmetry
proposed in
 ref.~\cite{rtv}:  the coset models
$\SU(2)_k \times \SU(2)_{l} \over \SU(2)_{k+l}$ with $l = n - {5 \over 2}$
and $k=1,\dots,n-1$.
The non unitary fluxes in the FSSG theory
correspond to  non-unitary fluxes between the models (\ref{f1},\ref{f2},
\ref{f3}).
For  $k=1$ we find the fluxes  $\M_{2n-1,2n+1} +\Phi_{13}\fr
\M_{2n-3,2n-1}+\Phi_{31}$ in agreement with the perturbative results obtained
 in ref.~\cite{flow},
this  seems to extend the validity of eqs.~(\ref{f1},\ref{f2},\ref{f3})
also for $l$,  as supported by the results in ref.~\cite{flow}.\\

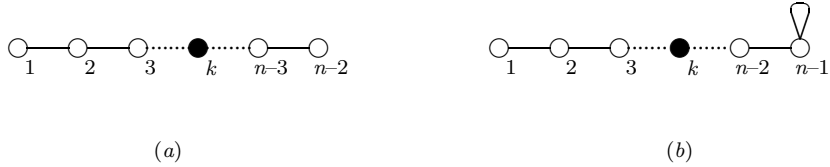
\begin{figure}[htbp]
\begin{center}
\begin{picture}(130,70)(0,-40)
\put(-10,25) {\usebox{\SAnTBA}}
\put(70,25){\usebox{\STNTBA}}
\put(25,13) {\makebox(0,0)[t]{{\protect\scr ({\em a})}}}
\put(110,13){\makebox(0,0)[t]{{\protect\scr ({\em b})}}}
\put(-20,-15){\parbox{130mm}{\caption{\label{fig13}\protect {\sts
(a)A diagram of the $\A_n(TBA)$ type is associated
with a quantum reductions of the FSSG.
In the case $k=1$, we get the $\M_{n}+\Phi_{13}$ minimal theories.
(b)Also the $\T_{n-1}(TBA)$ diagram is associated with a quantum reduction
of the
FSSG theory. The theories have
fractional supersymmetry.
In the $k=1$ case, we have the $\M_{2n-1,2n+1}+\Phi_{13}$ minimal theories.}}}}
\end{picture}
\end{center}
\end{figure}

\vspace{0.3cm}

\resection{ The theories  $ G_k \times G_l \over G_{k+l}$}
In this section we shall generalize the results obtained in {\em Section~1}
to  the
general coset theory  $G_k \times G_l \over G_{k+l}$
with $G$ any group generated
by the non-simply-laced $\G_2,\F_4,\B_n,\C_n$ algebras perturbed by the
operator
$\Phi^{id,id}_{adj}$. Let us begin with the theories
\eq
{(G_2)_1 \times (G_2)_1 \over (G_2)_2}+  \Phi^{id,id}_{adj}
\virg
\label{GGG}
\en
\eq
{(F_4)_1 \times (F_4)_1 \over (F_4)_2}+ \Phi^{id,id}_{adj}
\pu
\label{FFF}
\en
At the conformal point these are also particular theories of the minimal
series.
The theory (\ref{GGG}) is the model $\M_{9}$ perturbed by the relevant
operator $\Phi_{21}$ and  (\ref{FFF}) is the model  $\M_{10}$
perturbed by $\Phi_{12}$.
The scattering theories are those obtained by Smirnov in
ref.~\cite{sm1} using a quantum reduction of the Izergin-Korepin model.
The reduction must be realized at the point (see ref.~\cite{sm1}
for more details)
$ \tilde{\xi}= { 5 \pi \over 6}$
for the $\G_2$ model and
$ {\xi}= { 5 \pi \over 9}$
for  $\F_4$.
As expected in the  $\F_4$ case the mass spectrum consists in
four particles
(see {\em Table~\ref{tab1}}), and for   $\G_2$
 three of the four particles degenerate
so that the effective spectrum consists of two non degenerate particles.
The  mass ratio is
reported in {\em Table~\ref{tab1}}.
We do not go further into   this specific problem and
we address the  reader interested to these theories and the connection with the
Toda theory with parafermions to the work~\cite{tat}.
Here we mention that the counting
argument applied to these  models supports  the presence of conserved charges
with spin
$s=1,5$ for $\M_9$ and $s=1,5,7,11$
for $\M_{10}$. The above spins coincide  with  the
exponents of the  corresponding  Lie algebra.
These results and those  obtained in ref.~\cite{tat}
support
the following Y-system structure
for the $\G_2$ theory we have
\bea
Y_1 \lf( \th - { \imath \pi \over 4} \ri)
Y_1 \lf( \th + { \imath \pi \over 4} \ri) &=&
\lf(1+Y_{(2,3)}  \lf( \th + { \imath \pi \over 6} \ri)\ri)
\lf(1+Y_{(2,3)}  \lf( \th - { \imath \pi \over 6} \ri)\ri) \times \nn \\
&\times&
(1+Y_{(2,1)}( \th))~(1+Y_{(2,5)}( \th))~(1+Y_{(2,3)}( \th)) \times \nn \\
&\times&
\lf(1+Y_{(2,2)}  \lf( \th + { \imath \pi \over 12} \ri)\ri)
\lf(1+Y_{(2,2)}  \lf( \th - { \imath \pi \over 12} \ri)\ri) \times \nn\\
&\times&
\lf(1+Y_{(2,4)}  \lf( \th + { \imath \pi \over 12} \ri)\ri)
\lf(1+Y_{(2,4)}  \lf( \th - { \imath \pi \over 12} \ri)\ri) \times \nn
\eea
\bea
Y_{(2,3)} \lf( \th - { \imath \pi \over 12} \ri)
Y_{(2,3)} \lf( \th + { \imath \pi \over 12} \ri) &=&
(1+Y_{1}( \th)) \times \label{yy2}\\
&\times&
\lf(1+{1 \over Y_{(2,4)}( \th)}\ri)^{-1}
\lf(1+{1 \over Y_{(2,2)}( \th)}\ri)^{-1}
\nn
\eea
\bea
Y_{(2,k)} \lf( \th - { \imath \pi \over 12} \ri)
Y_{(2,k)} \lf( \th + { \imath \pi \over 12} \ri)&=&
\lf(1+{1 \over Y_{(2,{k+1})}( \th)}\ri)^{-1}
\lf(1+{1 \over Y_{(2,{k-1})}( \th)}\ri)^{-1} \nn
\eea
in eq. (\ref{yy2}) $k=1,2,4,5$. The nodes $1$ and $(2,3)$ are massive
while the others are magnonic.
For the $\F_4$ case we have
\bea
Y_1 \lf( \th - { \imath \pi \over 9} \ri)
Y_1 \lf( \th + { \imath \pi \over 9} \ri) &=& (1+Y_{2}(\th)) \nn \\
Y_2 \lf( \th - { \imath \pi \over 9} \ri)
Y_2 \lf( \th + { \imath \pi \over 9} \ri) &=& (1+Y_{1}(\th))
(1+Y_{(3,1)}(\th))(1+Y_{(3,3)}(\th)) \times \nn \\
&\times&
\lf(1+Y_{(3,2)}\lf(\th +{ \imath \pi \over 18} \ri)\ri)
\lf(1+Y_{(3,2)}\lf(\th -{ \imath \pi \over 18} \ri)\ri) \nn
\eea
\bea
Y_{(3,2)}\lf( \th - { \imath \pi \over 18} \ri)
Y_{(3,2)} \lf( \th + { \imath \pi \over 18} \ri) &=& (1+Y_{2}(\th))
(1+Y_{(4,2)}(\th)) \times \label{yy1}\\
&\times&
\lf(1+{1 \over Y_{(3,{1})}( \th)}\ri)^{-1}
\lf(1+{1 \over Y_{(3,{3})}( \th)}\ri)^{-1} \nn
\eea
\bea
Y_{(4,2)}\lf( \th - { \imath \pi \over 18} \ri)
Y_{(4,2)} \lf( \th + { \imath \pi \over 18} \ri) &=& (1+Y_{(3,2)}(\th))
\times \nn\\
&\times&
\lf(1+{1 \over Y_{(4,{1})}( \th)}\ri)^{-1}
\lf(1+{1 \over Y_{(4,{3})}( \th)}\ri)^{-1} \nn
\eea
\bea
Y_{(j,k)} \lf( \th - { \imath \pi \over 18} \ri)
Y_{(j,k)} \lf( \th + { \imath \pi \over 18} \ri) &=&
\lf(1+{1 \over Y_{(j,{2})}( \th)}\ri)^{-1}
(1+{ Y_{(j+1,{k})}( \th)}) \times \nn \\
(1+{ Y_{(j-1,{k})}( \th)})
\sps\sps j&=&2,3 \sps \sps k=1,3  \pu \nn
\eea
The nodes $i=1,2,(3,2),(4,2)$ are massive,  the others are
magnonic.
These
Y-systems defines the correct UV central charge
and, in agreement with the  conformal
dimension of the fields $\Phi_{12}$ and $\Phi_{21}$,
deal with
functions with periodicity  ${\tilde{h}+2 \over
\tilde{h}}$.
Finally the $IR$ asymptotic behaviour gives  for the mass ratio those
listed in
{\em Table \ref{tab1}} obtained from the Smirnov reduction.
The natural generalizations of the
Y-systems (\ref{yyy},\ref{yy2},\ref{yy1})
 are those proposed
in ref.~\cite{ku}
with a slightly different interpretation. Let us recall the structure of the
Y-system proposed by Kuniba and Nakanishi.
Let $\{ \alpha_i \}$ be the set of the simple roots of
a simple Lie Algebra of rank
$r$ and
define the Cartan matrix $C_{ab}$, a symmetrized  Cartan matrix $B_{ab}$
and the incidence  matrix $I_{ab}$ as follows
\eq
t_a= {2 \over \langle \alpha_a|\alpha_a\rangle}
\virg \sps
t_{ab}=max(t_a,t_b)
\en
\eq
C_{ab}=2 { \langle \alpha_a|\alpha_b\rangle  \over
\langle \alpha_a|\alpha_a\rangle }
\virg \sps
B_{ab}={t_b \over t_{ab}} C_{ab}
\virg \sps
I_{ab}=2 \delta_{ab}-B_{ab}
\pu
\en
The Kuniba-Nakanishi's Y-system is
\[
Y_{(a,m)} \lf(\th + { \imath \pi \over \tilde{h} t_a} \ri)
Y_{(a,m)} \lf(\th - { \imath \pi \over \tilde{h} t_a} \ri)=
{ \prod_{b=1}^{r} \prod_{k=1}^{3} F_k^{I_{ab}~\delta_{t_ak,t_{ab}}}
\over (1+Y_{(a,m-1)}(\th)^{-1})(1+Y_{(a,m+1)}(\th)^{-1})}
\]
\eq
F_k = \prod_{j=-k+1}^{k-1} \prod_{n=0}^{k-1-|j|} \lf( 1+ Y_{(b,t_b m /t_a +j)}
\lf(\th+{\imath \pi (k-1-|j|-2n) \over \tilde{h}t_b }\ri) \ri)
\label{kun}
\en
with
${ \tilde{h}=1}$, $Y(\th)=e^{-\varepsilon(\th)}$,
$\tilde{l}$ a positive integer, $\tilde{l}_a=t_a\tilde{l}$,
$1 \leq m \leq (\tilde{l}_a-1)$ and boundary
conditions $Y_{(a,0)}(\th)^{-1}=Y_{{(a,m)}}(\th)=0$ if $m \not\in Z$ (see
 the original work for more details).
They proposed this Y-system with all nodes massive
 for the description of the perturbation of
the generalized
Gepner parafermions $G/U(1)^r$. We propose to put in
(\ref{kun}) $\tilde{h}$ equal the
 dual Coxeter
number and $Y(\th)=e^{\varepsilon(\th)}$  with magnonic
asymptotic behaviour for all the nodes
but for the nodes $m=t_a k$ \sps $1 \leq k \leq (\tilde{l}-1)$ where
\eq
Y_{(a,m=t_a k)}(\th) \sim e^{Rm_a e^{|\th|}} \sps  R,|\th| \fr \infty
\pu
\label{ms}
\en
In eq. (\ref{ms}), the mass spectrum is as
usual the Perron Frobenius eigenvector if $G$ is generated by
an $\A\D\E$ algebra.
For $G$ generated by the
$\G_2,\F_4,\B_n,\C_n$ algebras the mass spectrum is listed in
{\em Table~\ref{tab1}}. Using the sum-rules cited in ref.~\cite{ku}
\eq
{ 6 \over \pi^2} \sum_{a~\in~all~nodes}  L \lf( { y_a
\over 1+ y_a} \ri) = { \tilde{l} dim G \over \tilde{l}+ \tilde{h}}-r
\en
and putting $k+l=\tilde{l}$, we find as  UV
central charge the value  of the
$G_k \times G_l \over G_{k+l}$   models~\cite{wtba}
also
in the non-simply-laced cases.
We  verified that the Y-systems  (\ref{kun}) define functions
with periodicity
$\tilde{h}+ \tilde{l} \over
\tilde{h}$, this is  in agreement with the  conformal dimension
of the field $\Phi^{id,id}_{adj}$. Substituting  to the massive term a right
mover ($m_a \cosh \th \fr m_a e^{\th} /2$) and putting a left mover
${m_a e^{-\th}/2}$ on the line $l=\tilde{l}-k$ we define
a massless flux   between the  theories
${G_k \times G_l \over G_{k+l}}$  and  $ {G_{l-k} \times G_k \over G_{l}}$.
In the IR region the attracting operator is the least irrelevant,
its
conformal dimensions are $h(\Phi_{act})=\bar{h}(\Phi_{act})=
1+ { \tilde{h} \over \tilde{h}+l-k}$
(see {\em Figure~\ref{fig23}}).
\section{CONCLUSIONS}
In this article we   analysed  the sine-Gordon
model for special values of the coupling constant where
 the theory is equivalent (up to an orbifold) to the coset
theories  $ SO(2n-1)_1 \times \SO(2n-1)_1 \over \SO(2n-1)_2$.
We found that at these points the S-matrix is not  diagonal. As a
consequence,
we must apply the Algebraic Bethe Ansatz technique to extract
information about the ground
state energy of the model. Despite  this complexity, the ABA works
in a rather simple
way. The result can be written in a universal form, and, using the
properties of the mass spectrum, we  obtained the associate Y-system and
verified
the correct periodicity and the correct UV behaviour.
In the second section  we
studied the
sine-Gordon model in special points of the repulsive regime related by a sort
of duality property to the theories  studied in the attractive regime.
In this part we also put forward  new remarks
 about the fractional supersymmetric sine-Gordon.
In the third part we have faced
the general problem of the
coset models ${G_k \times G_l \over G_{k+l}}+ \phi^{id,id}_{adj}$ generated by
a non-simply-laced Lie algebras.
We started the analysis
from the bottom models of the  $\G_2$ and $\F_4$  coset theory. The results
obtained for these models combined with those
obtained in the $\B_n$ case confirm a
Y-system structure generalizing  that proposed  by Kuniba and Nakanishi.
We hope to return to the topic soon  and to
complete it by  giving the set of S matrices for the non-simply-laced
Toda field theory  with parafermions and to confirm the S-matrix
obtained in ref.~\cite{des} for the $B_n$ cases.
\vskip 1cm
{\bf Acknowledgements}\\

We are greatly indebted  with
F.Gliozzi and F.Ravanini
for help and  useful discussions.
We thank  S.Moretti  for useful discussions.
\begin{figure}[h]
\begin{center}
\begin{picture}(130,170)(0,-20)
\put(0,120){\Large $\G_2:$}
\put(5,120){\usebox{\GG}}
\put(80,120){\usebox{\GGTBA}}
\put(70,120){\Large $\Rightarrow$}

\put(0,75){\Large $\F_4:$}
\put(5,75){\usebox{\FF}}
\put(80,75){\usebox{\FFTBA}}
\put(70,75){\Large $\Rightarrow$}

\put(0,45){\Large $\B_n:$}
\put(5,45){\usebox{\Bn}}
\put(80,45){\usebox{\BnTBA}}
\put(70,45){\Large $\Rightarrow$}

\put(0,15){\Large $\C_n:$}
\put(5,15) {\usebox{\Cn}}
\put(80,15){\usebox{\CnTBA}}
\put(70,15){\Large $\Rightarrow$}
\put(-20,-10){\parbox{130mm}{\caption{\label{fig22}\protect {\sts
$\G_2,\F_4,\B_n,\C_n$ diagrams. The numbers show the labelling of the
different nodes.
On the right of the arrows there is the Dynkin diagram, on the left
the corresponding TBA ``graph''
for the $G_1 \times G_1 \over G_2$ deformed coset theories.}}}}
\end{picture}
\end{center}
\end{figure}
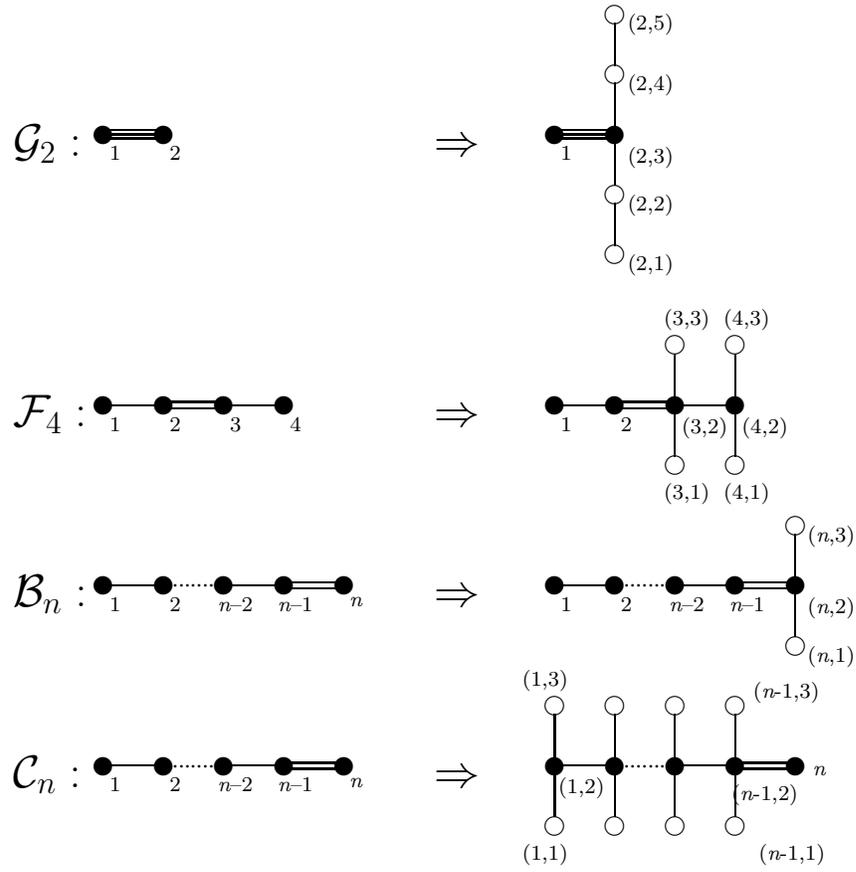
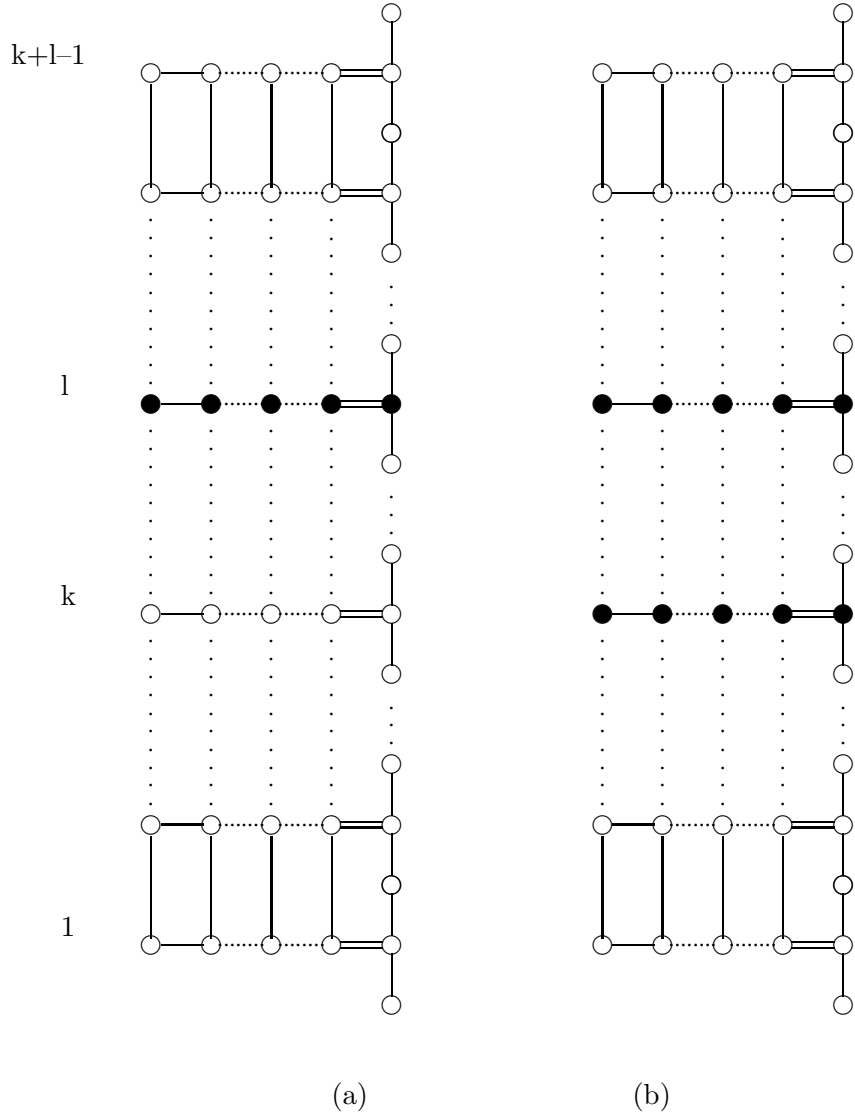
\begin{figure}[h]
\begin{center}
\begin{picture}(130,220)(0,-20)
\put(0,180){\usebox{\BBBTBA}}
\put(-15,185){ {\small  k+l--1}}
\multiput(10,164.5)(10,0){4}{\line(0,1){17}}
\put(0,160){\usebox{\BBBTBA}}
\multiput(50,142)(0,3){3}{\circle*{.25}}
\multiput(10,132)(10,0){4}{\multiput(0,0)(0,3){10}{\circle*{.25}}}
\put(0,125){\usebox{\BBBTBA}}
\multiput(10,128.5)(10,0){5}{\circle*{3}}
\put(-5,130){\small l}
\multiput(10,97)(10,0){4}{\multiput(0,0)(0,3){10}{\circle*{.25}}}
\multiput(50,107)(0,3){3}{\circle*{.25}}
\put(0,90){\usebox{\BBBTBA}}
\put(-5,95){{\small k}}
\multiput(10,62)(10,0){4}{\multiput(0,0)(0,3){10}{\circle*{.25}}}
\multiput(50,72)(0,3){3}{\circle*{.25}}
\put(0,55){\usebox{\BBBTBA}}
\multiput(10,39.5)(10,0){4}{\line(0,1){17}}
\put(0,35){\usebox{\BBBTBA}}
\put(-5,40){{\small 1}}
\put(75,180){\usebox{\BBBTBA}}
\multiput(85,164.5)(10,0){4}{\line(0,1){17}}
\put(75,160){\usebox{\BBBTBA}}
\multiput(125,142)(0,3){3}{\circle*{.25}}
\multiput(85,132)(10,0){4}{\multiput(0,0)(0,3){10}{\circle*{.25}}}
\put(75,125){\usebox{\BBBTBA}}
\multiput(85,128.5)(10,0){5}{\circle*{3}}
\multiput(85,97)(10,0){4}{\multiput(0,0)(0,3){10}{\circle*{.25}}}
\multiput(125,107)(0,3){3}{\circle*{.25}}
\put(75,90){\usebox{\BBBTBA}}
\multiput(85,93.5)(10,0){5}{\circle*{3}}
\multiput(85,62)(10,0){4}{\multiput(0,0)(0,3){10}{\circle*{.25}}}
\multiput(125,72)(0,3){3}{\circle*{.25}}
\put(75,55){\usebox{\BBBTBA}}
\multiput(85,39.5)(10,0){4}{\line(0,1){17}}
\put(75,35){\usebox{\BBBTBA}}
\put(40,12){{\small (a)}}
\put(90,12){{\small (b)}}
\put(-20,-10){\parbox{130mm}{\caption{\label{fig23}\protect {\sts
(a) The graph associated to the massive
${\SO(2n-1)_k \times \SO(2n-1)_l \over \SO(2n-1)_{k+l}}+\Phi^{id,id}_{adj}$
TBA systems. The white
nodes correspond to speudoparticles and the black to particles the mass ratio
is
in {\em Table~1}.
(b) The graph associated to the  massless
${\SO(2n-1)_k \times \SO(2n-1)_l \over \SO(2n-1)_{k+l}}+\Phi^{id,id}_{adj}$
TBA systems. The white
nodes correspond to speudoparticles and the black nodes correspond
to right or left movers
the scale ratio is
in {\em Table~1}. }}}}
\end{picture}
\end{center}
\end{figure}
\begin{table}[h]
\begin{center}
\begin{tabular}{||c|c|c|c||} \hline
\rll
{  $\G$}  &  Mass spectrum         &
 $\tilde{h}$      &    s    \\
\hline
\rl
{\large $\G_2$} &
$M_1=M$ ,
$M_2=2M \cos({\pi \over 12})$ &
4 &
1,5 \\
\hline
\rll {\large $\F_4$} &
$M_1=2M \cos({\pi \over 18})$ ,
$M_2=4M \cos({\pi \over 18})\cos({\pi \over 9})$ &  9 &
1,5,7,11\\
\rl  & $M_3=4M \cos({\pi \over 18})\sin({2 \pi \over 9})$,
$M_4=2M \sin({2 \pi \over 9})$ & & \\
\hline
\rl {\large $\B_n$} &
$M_n=M$ , $ M_i = 2 M\sin\lf(  { \pi i \over  \tilde{h}} \ri)$,\sps i=1,2,\dots
,$n$-1 &
$2n-1$ &
1,3\dots,2$n$-1\\
\hline
\rl
{\large $\C_n$} &
$M_i=2 M \sin\lf( {\pi i \over 2 \tilde{h}} \ri)$,\sps i=1,2,\dots ,$n$ &
$n$+1 &
1,3\dots,2$n$-1\\
\hline
\end{tabular}
\parbox{130mm}{\caption{\label{tab1}}
\protect{\sts Mass spectrum,  Coxeter numbers $\tilde{h}$ and
Coxeter exponents $s$. }}
\end{center}
\end{table}
\begin{table}[h]
\begin{center}
\begin{tabular}{||c|c||} \hline \hline
\rl
 {\large $\G_2$} &
 $ 2 \cos\lf( { \pi \over \tilde{h}} \ri)M_1 = M_2
+2 \cos\lf( {2 \pi \over 3
 \tilde{h} } \ri)M_2$~,~
 $2 \cos\lf( {\pi \over 3 \tilde{h}} \ri)M_2 = M_1$ \\ \hline
\rl
{\large  $\F_4$ } &
\rl
$ 2 \cos\lf( { \pi \over 2 \tilde{h}} \ri) M_4= M_3$~,~
$ 2 \cos\lf( { \pi \over 2 \tilde{h}} \ri) M_3= M_4 + M_2$ \\
\rl &
$ 2 \cos\lf( { \pi \over  \tilde{h}} \ri) M_2= M_1+
 2 \cos\lf( { \pi \over 2 \tilde{h}} \ri) M_3$~,~
$ 2 \cos\lf( { \pi \over  \tilde{h}} \ri) M_1= M_2$ \\
\hline
\rl
{\large $\B_n$} &
\rl
$2 \cos\lf( { \pi \over \tilde{h}} \ri) M_i = M_{i-1} + M_{i+1}\virg
{}~i \le n-2$ \\
\rl &
$2 \cos\lf( { \pi \over \tilde{h}} \ri) M_{n-1} = M_{n-2} +2 \cos\lf( { \pi
\over 2 \tilde{h}}\ri) M_n  $~,~
$ 2 \cos\lf( { \pi \over 2 \tilde{h}} \ri)M_n=M_{n-1}$ \\
\hline
\rl
{\large $\C_n$} &
\rl
$2 \cos\lf( { \pi \over 2 \tilde{h}} \ri) M_i = M_{i-1} + M_{i+1}\virg~i
\le n-1$
\\
\rl &
$2 \cos\lf( { \pi \over \tilde{h}} \ri) M_n = 2\cos\lf( { \pi
\over 2 \tilde{h}}\ri) M_{n-1}  $ \\
\hline \hline
\end{tabular}
\parbox{130mm}{\caption{\label{tab2}}
\protect{\footnotesize  Mass properties in non-simply-laced $G_k \times G_l
\over G_{k+l}$ deformed coset theories.} }
\end{center}
\end{table}

\end{document}